\documentclass{aa}  
\usepackage{graphicx}
\usepackage{txfonts}
\newcommand{\allframe}{{\small ALLFRAME}}
\begin{document}
   \title{The Peculiar Horizontal Branch Morphology of the Galactic Globular Clusters NGC~6388 and NGC~6441: new insights from UV observations}

   \author{G. Busso  \inst{1,2,3},
          S. Cassisi \inst{1}, 
          G. Piotto \inst{3},
	  M. Castellani \inst{4},
          M. Romaniello \inst{5},
          M. Catelan \inst{6}, 
          S.~G. Djorgovski \inst{7},
          A. Recio Blanco \inst{8}, 
          A. Renzini \inst{9},
          M.~R. Rich \inst{10},
          A.~V. Sweigart \inst{11}
          \and
          M. Zoccali \inst{6}
          }

   \institute{ INAF - Osservatorio Astronomico di Collurania, via M. Maggini, 64100 Teramo, Italy\\   
   \email{busso@oa-teramo.inaf.it, cassisi@oa-teramo.inaf.it}
   \and Institut f\"ur Theoretische Physik und Astrophysik,       
   University of Kiel, Leibnizstrasse 15, 24114, Kiel, Germany \\      
   \and Dipartimento di Astronomia, Universit\`a di  Padova, Vicolo  
   dell'Osservatorio  2, I-35122 Padova, Italy \\
   \email{giampaolo.piotto@unipd.it}
   \and INAF - Osservatorio Astronomico di Roma, Via di Frascati 33, 00040, Monte Porzio Catone, Roma, Italy \\      
   \email{m.castellani@mporzio.astro.it}
   \and European Southern Observatory, Karl Schwarzchild Strasse 2, 85748, Garching, Germany \\
   \email{mromanie@eso.org}    
   \and Pontificia Universidad Cat\'olica de Chile, Departamento de
       Astronom\'\i a y Astrof\'\i sica, Av. Vicu\~{n}a Mackenna 4860,
       782-0436 Macul, Santiago, Chile \\   
   \email{mcatelan@astro.puc.cl, mzoccali@astro.puc.cl}  
   \and Astronomy Department, California Institute of Technology, MC 105-24, Pasadena, CA 91125, USA \\   
   \email{george@astro.caltech.edu}
   \and Observatoire de la C\^ote d®Azur, Dpt. Cassiop\'ee, CNRS/UMR6202,
 B.P. 4229, 06304 \\    
   \email{arecio@obs-nice.fr} 
   \and INAF-Osservatorio Astronomico di Padova, vicolo Osservatorio 5, I-35122 Padova, Italy \\  
   \email{alvio.renzini@oapd.inaf.it} 
   \and Department of Physics and Astronomy, UCLA, 430 Portola Plaza, Los Angeles CA, 90095-1547, USA \\     
   \email{rmr@astro.ucla.edu}
   \and NASA Goddard Space Flight Center, Code 667, Greenbelt, MD 20771, USA \\  
   \email{allen.v.sweigart@nasa.gov} 
   }

   \date{}

  \abstract
{In this paper we present multiband optical and UV Hubble Space
Telescope photometry of the two Galactic globular clusters NGC~6388
and NGC~6441.} 
{We investigate the properties of their anomalous horizontal branches 
in different photometric planes in order to shed light on the nature
of the physical mechanism(s) responsible for the
 existence of an
extended blue tail and of a slope in the horizontal branch, visible in all the
color-magnitude diagrams.} 
{New photometric data have been collected and carefully
reduced. Empirical data have been compared with updated stellar models
of low-mass, metal-rich, He-burning structures, transformed to the
observational plane with appropriate model atmospheres.}
{We have obtained the first UV color-magnitude diagrams for NGC~6388
and NGC~6441. These diagrams confirm previous results, obtained in
optical bands, about the presence of a sizeable stellar population of
extremely hot horizontal branch stars. At least in NGC~6388, we find a
clear indication that at the hot end of the horizontal branch the
distribution of stars forms a hook-like feature, closely resembling
those observed in NGC~2808 and $\omega$ Centauri. We briefly review
the theoretical scenarios that have been suggested for interpreting
this observational feature. We also investigate the tilted
horizontal branch morphology and provide further evidence that
supports early suggestions that this feature cannot be
interpreted as an effect of differential reddening. We show
that a possible solution of the puzzle is to assume that a
small fraction - ranging between 10-20$\%$ - of the
stellar population in the two clusters is strongly helium enriched
(Y$\sim$0.40 in NGC~6388 and Y$\sim$0.35 in NGC~6441). 
The occurrence of a spread in the He abundance between the canonical value (Y$\sim0.26$) and
the quoted upper limits can significantly help in explaining the \lq{whole}\rq\ morphology
of the horizontal branch and the pulsational properties of the variable stars in the target clusters.}
   {}

\keywords{stars: horizontal-branch -- ultraviolet: stars -- globular cluster: individual (NGC~6388, NGC~6441)}

\authorrunning{Busso et al.}
\titlerunning{The peculiar Horizontal Branch morphology of NGC~6388 and NGC~6441} 
\maketitle


\section{Introduction}

The study of galactic globular clusters (GCs) is of paramount
relevance in order to understand the process of formation of the Milky Way
as well as for its implication for many topics related to stellar evolution theory and,
in particular for the investigation of the evolutionary properties of low-mass stars.

Even if we are able in general to provide a detailed interpretation of the various empirical sequences observed in the 
color-magnitude diagrams (CMDs) of Galactic GCs in terms of the evolutionary and structural properties of 
both H- and He-burning low-mass stars, many important issues still wait for a clear comprehension.
One of the most relevant issues is related to the long-standing problem of the $2^{nd}$ parameter concerning the morphology of the horizontal branch (HB) (see \cite{catelan07} and references therein). In these past few years, it is becoming more and more evident that we need to revise our point of view concerning the \lq{nature}\rq\ of $2^{nd}$ parameter: there is  increasing evidence that different $2^{nd}$ parameters could be at work in different GCs, as well as that more than a \lq{single}\rq\ parameter - besides metallicity (the first parameter) - could be at work in the same cluster (e.g., Lee et al. 2005 and references therein).
In this context, a strong contribution has been provided  by the observational data for the two metal-rich ([Fe/H]$\approx-0.5$) 
bulge clusters NGC~6388 and NGC~6441 collected by Rich et al.~(1997) as part of a Hubble Space Telescope (HST) project.
The optical photometric data collected by Rich et al. disclosed a number of important and unexpected empirical facts:

\begin{itemize}

\item both clusters show not only the red HB, characteristic of metal-rich GCs as 47~Tuc ([Fe/H]$=-0.76$,  Harris 1996), 
but also a quite extended blue HB. 
The effective temperature
($T_{\rm eff}$) reached by the blue HB stars
  in both clusters is so high that their F555W magnitudes are
  dominated by the huge bolometric corrections, and the (F439W-F555W)
  color index is no more a good $T_{\rm eff}$ indicator. As a
  consequence, the blue HB (BHB) appears as a vertical branch, this
  occurrence hampering a more detailed investigation of the hot HB 
  stellar populations in these clusters. The fact that NGC~6388 and
  NGC~6441 show not only the \lq{canonical}\rq\ stubby red HB, but
  also an extended blue tail, was the first sound evidence that the
  second parameter is also at work in old, metal-rich stellar systems;

\item even more surprisingly, the CMD of Rich et al.~(1997)
  clearly shows that the mean HB brightness at the top of
  the blue HB tail is roughly 0.5 magnitudes brighter in the F555W band than
  the red HB portion, which appears significantly sloped as well. 
  This slope, as pointed out in \cite{sweigart98} is not present in the theoretical models, and is difficult
  to explain (see also the discussion in Raimondo et al. 2002). 

\end{itemize}

In an attempt to interpret the occurrence of the blue tail along the
HB as well as of its tilted morphology, many non-canonical scenarios
and/or observational effects have been suggested (e.g. \cite{sweigart98}, \cite{raimondo02}, \cite{moehler06b}, \cite{caloi06}; see also \cite{catelan07} for a review). We will enter in more details into this topic 
in the following discussion.

In the last few years, a large effort has also been devoted to the
investigation of the properties of the population of variable stars in 
both clusters (e.g. \cite{pritzl02}, \cite{clementini05}, \cite{corwin06}, \cite{matsunaga07}).  
An important and unexpected result - somehow related to the
peculiar HB morphology of NGC~6388 and NGC~6441 - was the evidence that
their RR Lyrae populations do not behave as in typical Oosterhoff type
I systems, as should have been expected on the basis of their metal
content. On the contrary, 
these analyses have shown the presence of a significant fraction of long-period c-type RR Lyrae, which are
not commonly seen in most other globular clusters (\cite{catelan04} and references therein),
while in NGC~6388, but not in NGC~6441, some short-period c-type RR-Lyrae 
(i.e., with $P \leq 0.3$~d) have been observed.

According to Layden et al.~(1999) and Pritzl et al.~(2002), the morphology of the
light curves of the c-type RR Lyrae in both clusters seems also to be
peculiar when compared with that of the first-overtone RR Lyrae
observed in other clusters, which must be due to peculiar values of some 
physical parameters (such as chemical composition, envelope mass). 
In addition to harboring this anomalous population of RR
Lyrae, NGC~6388 and NGC~6441 contain a significant number of type II
Cepheids, making them the most metal-rich clusters known to contain
this kind of variables.

In summary, it is clear that NGC~6388 and NGC~6441 seem to be very
unusual clusters in many respects, and their observational
characteristics, in particular the photometric properties, and  their RR Lyrae populations severely challenge both the evolutionary and pulsational models of low-mass, metal-rich stellar
structures.

With the aim of better understanding the observational properties of the
hot HB stars belonging to these clusters, in this work we present the
results of an HST project (GO8718, PI: G. Piotto) devoted to collect
HST WFPC2 multiband photometry in the filters: F255W, F336W, F439W and
F555W.  
The main parameters of NGC~6388 and NGC~6441, as given
by  Harris (1996, in the 2003 revision), are summarized in Table 1.

\begin{table}
\begin{center}
\caption{
Fundamental parameters of NGC~6388 and NGC~6441\protect\footnotemark{}.}
\begin{tabular}{| c | c | c |}
\hline
   & NGC~ 6388 & NGC~6441 \\
\hline
$l$  & 345.56 & -353.53 \\
\hline 
$b$  & -6.74 & -5.01 \\
\hline
$d_{\odot}$ & 10.0 & 11.7 \\
\hline
[Fe/H] & -0.60 & -0.53 \\
\hline
 $E(B-V)$ & 0.37 & 0.47 \\
\hline 
\end{tabular}
\end{center}
\end{table}

\footnotetext{l and b are respectively galactic longitude and latitude (in degrees), $d_{\odot}$ is the
distance from the Sun (in Kpc), [Fe/H] is the metallicity, and
$E(B-V)$ is the reddening (from the latest version of the Harris 1996
catalog).}.

 The data presented in this paper represent an extension of the
original F439W-F555W vs. F555W CMDs
published by \cite{rich97} and Piotto et al.~(2002). 
Because of the strong bolometric correction and temperature
insensitivity, the optical data are not as useful in the study of the
hottest HB stars, which can be better studied with the UV observations
discussed in the next sections.


\section{Observations and data reduction}  

\begin{table*}[t]
\centering
\begin{tabular}{| l | c | c | c | c |}
\hline
Cluster & Filter & number images & $t_{exp}$ [s] & Program ID \\
\hline
\bf{NGC~ 6388} & F255W & 7 & $1\times 1100, 3\times 1300, 3\times 1400$ & GO8718 \\
 & F336W & 3 & $1\times 260, 2\times 400$ & GO8718 \\
 & F439W & 3 & $1\times 50, 2\times 160$ & GO6095 \\
 & F555W & 2 & $1\times 12, 1\times 50$ & GO6095 \\
\hline
\bf{NGC~ 6441} & F255W & 7 & $1\times 1100, 3\times 1300, 3\times 1400$ & GO8718 \\
 & F336W & 3 & $1\times 260, 2\times 400 $ & GO8718 \\
 & F439W & 3 & $1\times 50, 2\times 160 $ & GO6095 \\
 & F555W & 2 & $1\times 14, 1\times 50 $ & GO6095 \\
\hline
\end{tabular}
\caption{Log of the observations}
\end{table*} 

Table 2 summarizes the most relevant information on the
observational data set used in the present paper.  The data come from
HST/WFPC2 observations; in all cases, the PC camera was centered on the
cluster center. The images were processed following the recipe in
\cite{silbermann}: the vignetted pixels and bad pixels and columns
were masked out using a vignetting frame created by P.~B. Stetson,
together with the appropriate data-quality file for each
frame. The single chip frames were extracted from the
4-chip-stack files and analyzed separately.

The photometric reduction was carried out using the {\small
DAOPHOT}II/{\small ALLFRAME} package (\cite{stetson87}, 1994). 
The F439W and F555W data were reduced within the
snapshot project (for more details, see \cite{piotto02}).  As for the
F255W and F336W data, for each image we calculated the appropriate
Point Spread Function (PSF).  The stars used to calculate the PSF
were selected carefully, as in the UV images, especially in the F255W
band, where many cosmic rays were present.
  
With {\small ALLSTAR}, we fitted on each image the best PSF, obtaining
an approximate list of stars for each single frame. This list was used
to match the different frames accurately and to find the correct
coordinate transformations among the frames.

{\allframe} needs also an input star list. We ran this program with
two star lists: the first one was obtained 
from the median image coming from all the F255W images, and the second
one was obtained on the median image  of the F336W frames. 

For each of the WFPC2 chips, a catalog of mean magnitudes was created
for both F255W and F336W bands; for the photometry in the F439W and F555W
bands, we used the HST snapshot data (Piotto et al.~2002).
The four photometries were finally combined in order to obtain the various
CMDs.

\subsection{The photometric calibration}    

In order to calibrate our photometry, we followed the procedure
outlined by \cite{dolphin00}, which allows also 
to correct the observed magnitudes for charge-transfer efficiency
effects (CTE). After having measured the zero point differences between
the {\allframe} magnitudes and the reference aperture of 0.5~arcsec, we applied
the following calibration equation to transform the magnitudes into
the HST Flight Photometric System (STMAG): 

\begin{center}
\begin{equation}
\textrm{WFPC2} =-2.5 \log(\textrm{DN}~\textrm{s}^{-1})+ Z_{FG} + \Delta Z_{FG} - \textrm{CTE}, 
\end{equation}
\end{center}

\noindent 
where DN is the flux in digital counts, CTE is the CTE correction, $Z_{FG}$ is the zero point, different
for each filter, and $\Delta Z_{FG}$ is the zero-point modification
for chip and gain settings.  Since the F255W zero point
is available only for the WFPC2 Synthetic System (Holtzman et
al.~1995), for consistency we have used the synthetic value also for the F336W filter.

\section{The optical and UV CMDs and star counts}\label{sec:cmds_counts}

\subsection{The color-magnitude diagrams}

\begin{figure*}
\centering
\includegraphics[width=17cm,height=18cm]{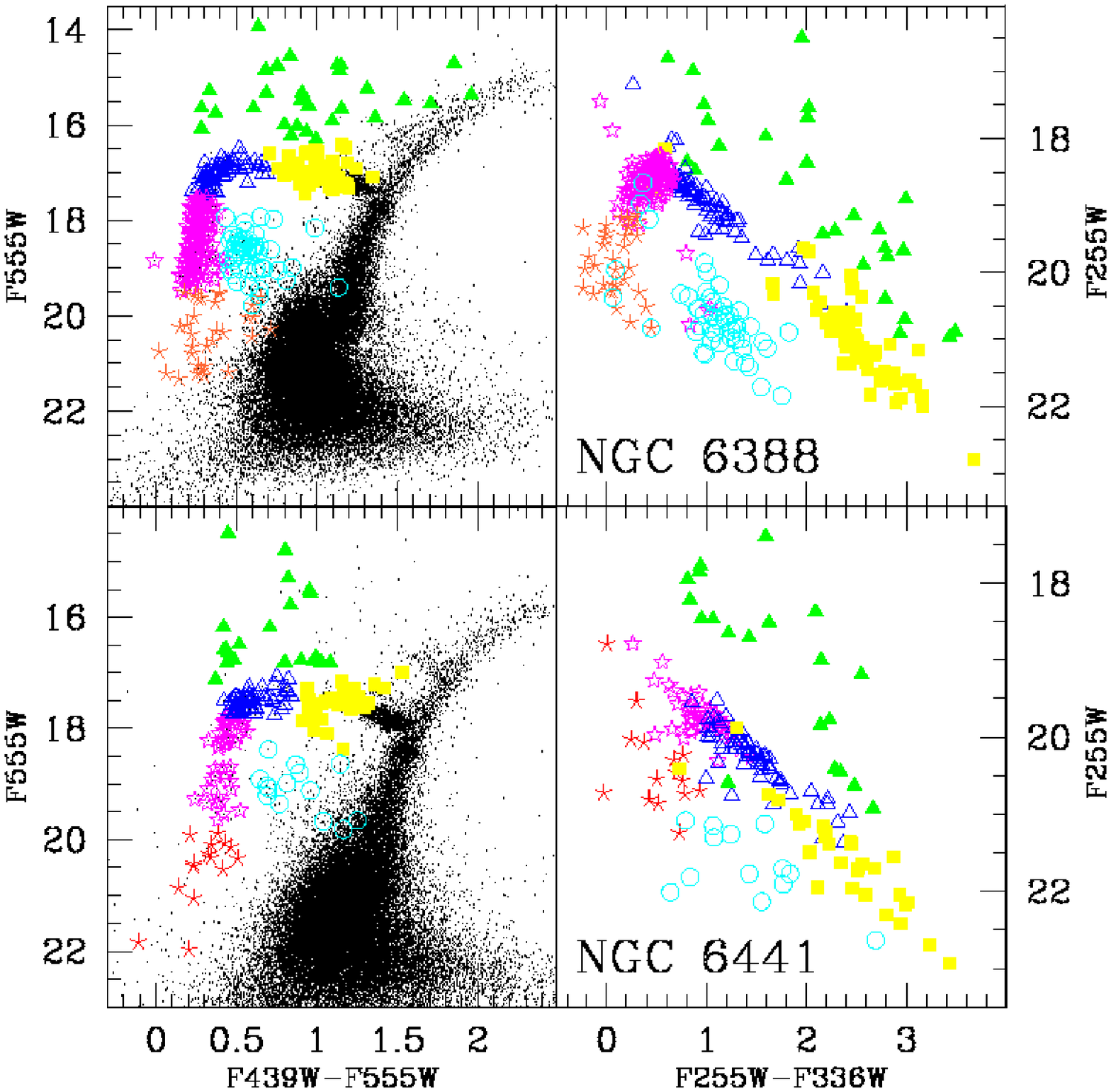}
\caption{Optical and UV CMDs for NGC~6388 (upper panels) and NGC~6441
(lower panels).  The different symbols and colors refer to the same
group of stars appearing both in the optical and UV CMD: yellow-filled squares indicate the red HB, 
blue open triangles the blue HB, magenta open stars the extended blue HB tail, the filled green triangles 
the post HB, the cyan open circle the blue stragglers candidates, and the orange asterisks are the blue 
hook candidates.}
\label{fig:box}
\end{figure*}

In Fig.~\ref{fig:box} we show the optical and UV band CMDs for both
clusters. Already a quick look at the two optical CMDs clearly shows
the most relevant, common properties of these two clusters, i.e., the
occurrence of an extended HB, and the tilt of the HB, with the HB on
the blue side of the instability strip brighter than the branch on the
red side. However, the HB morphology in
the optical CMDs differs in the two clusters in a number of significant
aspects:

\begin{itemize}

\item The HB on the blue side of the instability strip appears more
  populated in NGC~6388 than in NGC~6441.  This is not due to a
  sampling effect, since the number of sampled stars is very
  similar for the two clusters, as discussed below.  

\item The blue tail of NGC~6388 seems more extended than in NGC~6441. In
NGC~6388 the blue tail extends to more than 1 magnitude fainter than
the cluster turn-off (down to $F555W\approx 21.5$), whereas in the
case of NGC~6441 it barely reaches  the turn-off brightness,
at $F555W\approx 20.5$. We note that in this analysis
we do not consider the two stars at $F555W\approx 22$ as cluster members since
they are quite separated from the blue HB tail. 
Very likely, they do not belong 
to the cluster but instead to the bulge field population (hot HB stars have been found also in other 
bulge fields, see Busso et al.~2005); 

\item In the CMD of NGC~6441 there is a gap at $F555W\approx 18.5$,
  about one magnitude below the HB turning down; no such gap is
  visible in the HB of NGC~6388, which, instead seems to show a gap at
  $F555W\approx 19.8$. The latter is not present in NGC~6441, but this
  could be due to the smaller number of stars present in that region of
  its HB combined with photometric errors, to the lower sensibility
  to the temperature of the F255W-F336W color index, as well as to the larger differential reddening
  affecting this cluster (see discussion in Section 4.2). No such gaps are seen in the UV CMDs (see below). 
  However, one should also bear in mind that the occurrence of gaps could be simply due to statistical fluctuations
  (see the discussion in Catelan et al.~1998).

\end{itemize}

The right panels of Fig.~\ref{fig:box} show the UV CMDs for both clusters. In
order to make easier the cross-identification of the same sequences in
the optical and UV CMDs, we use different symbols and colors for
different stellar groups. In the UV photometry, only the hottest stars
are sufficiently bright to be properly measured, namely the blue
stragglers (BS), the bluer HB (BHB) stars, the Extreme Horizontal Branch (EHB) stars and the post-HB stars.  
The coolest stars
of the red HB clump are totally missed. It is also worth noting how
well the BS sequence is separated from the blue HB tail in the UV
CMD. This fact is more evident for NGC~6388 than for NGC~6441, possibly
as a consequence of the larger differential
reddening which characterizes NGC~6441 (see discussion in Section~\ref{sec:diffredd}) and the seemingly much larger BS population in the former.

Also the UV CMDs for both clusters show some noteworthy features: 

\begin{itemize}

\item In both CMDs the brightest
stars in the F255W band are not the stars populating the hottest
portion of the blue tail in the optical CMD;

\item The gaps visible in the optical bands
are not clearly visible in the UV CMDs,
making it impossible to establish whether these gaps are real features (as
clearly seen in NGC~2808, \cite{sosin97}, \cite{bedin00}), or 
simply a statistical fluctuation in the distribution in temperature
along the HB (Catelan et al.~1998). The only possible exception is the abrupt fall down in
the NGC~6388 HB star counts at F255W$\approx 19.4$, which corresponds to the gap at 
F555W$\approx 19.8$ in the optical CMD. This feature is relevant for
the discussion in Sec.~\ref{sec:bluehook}.

\item There is a large color spread at the hot end of the observed HB
sequence, of the order of $\sim0.5$ magnitudes in $F255W-F336W$,
larger than the photometric errors (less than 0.1 magnitudes).

\end{itemize}

\subsection{Star counts}

\begin{table*}[htdp]
\begin{center}
\begin{tabular}{|c|c|c|c|c|c|c|}
\hline
                   & $N_{\rm RGB}$ &  $N_{\rm HB}$ & $N_{\rm rHB}$ & $N_{\rm vHB}$ & $N_{\rm bHB}$ & $N_{\rm gHB}$ \\         
\hline
NGC~6441              & 705 &  1289 & 1122  & 21 & 146 & 61 \\
NGC~6388              & 664 &  1293 & 1056  & 19 & 218 & 154\\
ratio(6441/6388)     &  $1.06\pm 0.06$  &  $1.00 \pm 0.04$ & $1.06\pm 0.05$ & $1.11\pm 0.35$  & $0.70\pm 0.07$  &  $0.40\pm 0.06$ \\
\hline
\end{tabular}
\end{center}
\caption{Star counts for selected regions of the CMDs of the two clusters (see text for more details).}
\end{table*}

Table~3 lists the star counts in a number of relevant parts of the CMD
for the two clusters. $N_{\rm RG}$ is the number of 
red giant  (RG) stars\footnote{It accounts for RGB stars as well as for Asymptotic Giant Branch (AGB)  ones. We note that, due to the presence of differential reddening, it would be impossible to distinguish between RGB stars and AGB ones. However, for the present investigation, this is
not a relevant problem. } from
the magnitude level corresponding to the average magnitude of the red
HB to the tip of the Red Giant Branch (RGB), $N_{\rm HB}$ is the total number of HB stars,
$N_{\rm rHB}$ is the number of HB stars redder than the instability
strip, $N_{\rm vHB}$ the number of HB stars within the instability strip (i.e. RR Lyrae stars),
$N_{\rm bHB}$ the number of stars bluer than the instability strip,
$N_{\rm gHB}$ the number of stars bluer than the Grundahl et al.~(1999)
jump, at $T_{\rm eff} \sim 11,\!500$~K. The HB and RGB have been extracted from the (F336W-F555W,
F555W) CMD, as in this diagram it is easier to separate
the different CMD parts (see the left panels of Fig.~2), while the star
counts $N_{\rm rHB}$, $N_{\rm vHB}$ and $N_{\rm bHB}$ have been
extracted from the (F439W-F555W, F555W) diagram (see the right 
panels of Fig.~2). 

These star counts are surely
affected by incompleteness. However, as the two clusters have rather
similar HB morphologies, comparable apparent distance moduli, and the
observational data have been collected and reduced in the same way, we
expect, as a first approximation, similar parts of their CMDs to be
affected by similar incompleteness levels.  Now, the ratio of the
number of RGB stars 
in NGC 6441 over the number of RGB stars in NGC 6388 ($1.06\pm0.06$)
tells us that we are sampling approximately the same total number of
stars in the two clusters. Consistently, also the ratio of the total
number of HB stars ($1.00\pm0.04$) is comparable.  Interestingly
enough, the relative number of red HB stars is the same for the two
clusters (ratio $1.06\pm0.06$).

The relative number of stars within the instability strip is not very
informative (too few stars), and it must be noted that this is also
the region with the highest contamination by field stars. In any case,
we note that NGC~6388 has 22 known RR Lyrae, while
NGC~6441 has 68 RR Lyrae (\cite{corwin06}), confirming an
overabundance of RR Lyrae in the latter.  Most interestingly, as
suspected from the first visual impression, the blue HB in NGC 6441 is
much less populated than in NGC~6388, and the lack of hot stars in
NGC~6441 with respect to NGC~6388 is even more pronounced for $T_{\rm
eff}\ge11,500$K, i.e. beyond the Grundahl et al. (1999) jump.  Only
$11\pm1$\% of the HB stars of NGC~6441 populate the HB on the blue
side of the instability strip, to be compared with the $17\pm1$\% of
blue HB stars in NGC~6388 
(the ratio is 0.70$\pm$0.07). NGC~6441 has only $4.7\pm0.6$\% of its
stars hotter than the Grundahl et al. jump, while
$11.9\pm1.0$\% of the NGC~6388 HB stars are beyond the jump 
(the ratio is 0.40$\pm$0.06).  Whatever the
mechanism responsible for the anomalous HB, it apparently acts to a
different extent in the two clusters.


\begin{figure}
\begin{center}
\includegraphics[width=9cm,height=9cm]{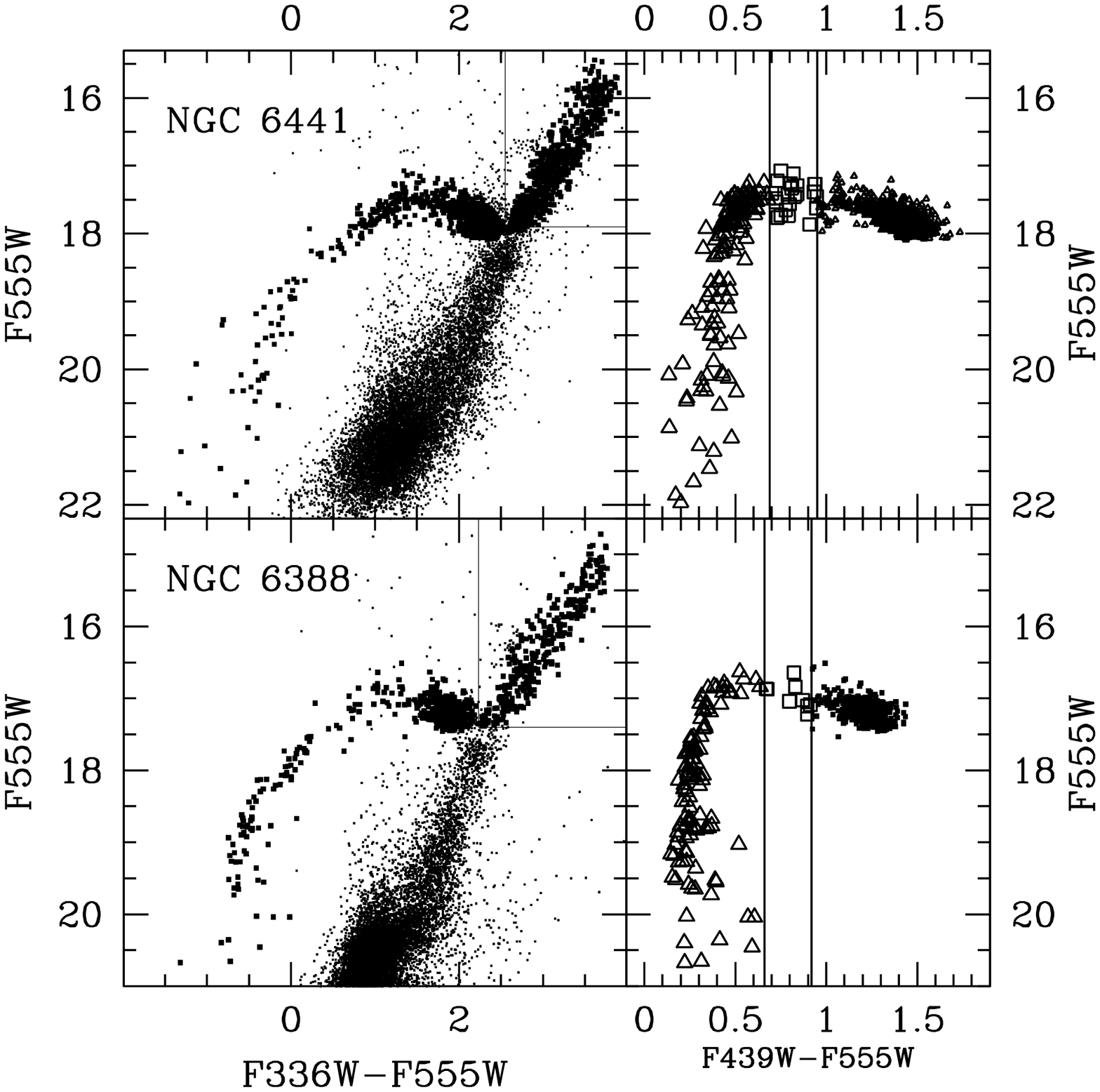}
\caption{In the left panels, the (F336W-F555W,F555W) diagrams for both
clusters, where the RGB (in the upper right box) and HB selection are
shown marked in black. In the right panels, for both clusters the
selection in the (F439W-F555W, F555W) for red (points), variable (open
squares), and blue (open triangles) HB stars is shown. The two vertical lines
represent the RR Lyrae instability strip boundaries ($0.21<$~F439W-F555W~$<0.47$),
from Bono et al.~(1997), reddened with the appropriate reddening for each cluster.}
\end{center}
\label{fig:sel}
\end{figure}

\section{The anomalous HBs of NGC~6388 and NGC~6441}

As extensively discussed in the introduction, the HBs of both clusters
in the optical CMDs show some relevant features that make them
extremely peculiar,
i.e. the extension of the HB to very hot temperature and its tilt.

So far, both effects have not been given a firm interpretation (see Catelan 2007 for a recent review). In this work, we try to take advantage of the
availability of multi-band photometric data in order to provide additional pieces of
information useful for constraining the evolutionary scenario.

In order to perform a detailed comparison between theory and
observations, we have adopted the Zero Age Horizontal Branch (ZAHB) models provided by Pietrinferni
et al.~(2006), supplemented by additional computations performed for
this specific project. We have adopted a metallicity Z=0.008 and He
contents Y=0.256.  All of the He-burning models used for the present work
correspond to an RGB progenitor whose age at the RGB tip is equal to
about 13Gyr. 
More details about these models\footnote{The complete set of evolutionary models can be downloaded 
from the following URL site: http://www.oa-teramo.inaf.it/BASTI/index.php.}
as well as about the adopted physical inputs and numerical assumptions can be found in
Pietrinferni et al.~(2004, 2006). 

Bolometric magnitudes and effective temperatures have been transformed
into HST magnitudes and colors according to transformations provided by
\cite{origlia00} based on the atmosphere models computed by \cite{bessel98}.

\subsection{Effects of the reddening in UV bands}\label{sec:reddUV}

When comparing photometric data with theoretical models, it is necessary
to account for reddening and extinction effects. In this context,
it is worth noting that both NGC~6388 and NGC~6441 are affected by substantial interstellar extinction 
(see data in Table 1). As a consequence, an
appropriate correction for reddening has to be applied before
comparing models with observations. 

It is a common procedure to adopt
an average extinction law for all stars, regardless of their spectral
type. However, the reddening correction depends on the stellar
effective temperature. While the commonly adopted approach is surely
suitable for stellar systems with a low extinction
[$E(B-V)\le0.10-0.15$], it is inappropriate for heavily reddened
objects, and the problems increase for blue/UV photometric bands (Grebel \& Roberts~1995).

It is evident that the presence of a large amount of interstellar extinction has no
relation at all with the presence of an extended blue tail along the HB of both clusters. However,
it could at least partially help explain the other striking feature observed in the HBs, i.e.
their tilted morphology.


\begin{figure}
\begin{center}
\includegraphics[width=9cm,height=9cm]{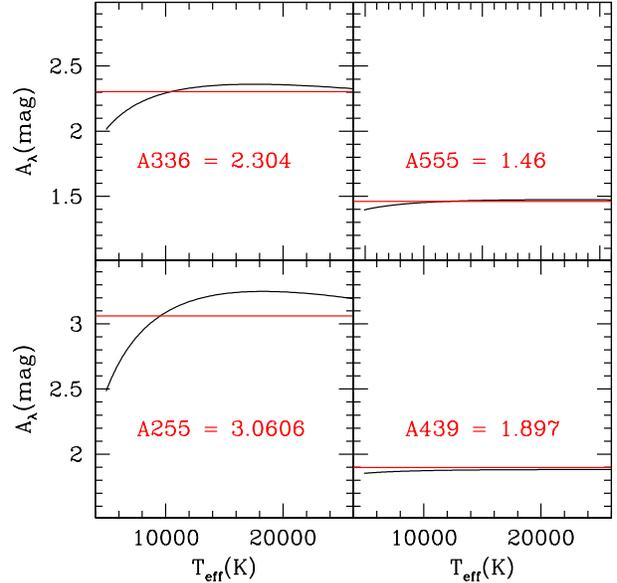}
\caption{The extinction coefficients are plotted as a
function of the effective temperature for the various photometric
bands used in the present work, assuming $E(B-V)=$0.45. The horizontal line shows the average
extinction coefficients for the same bands as provided by \cite{holtzman95}. }
\end{center}
\label{fig:redd}
\end{figure}


In order to adopt an accurate reddening correction, we have determined
its dependence on $T_{\rm eff}$ by convolving the appropriate model
atmosphere of \cite{bessel98} with the filter bandpass, and applying
the extinction law of \cite{savage}. Fig.~3 shows the size of this
effect. For the filters F555W and F439W, the extinction correction
marginally depends on the stellar $T_ {\rm eff}$, and the average
value provided by the table in \cite{holtzman95} can be safely used to
correct for extinction in these bands. For the bluer filters
the difference is larger, and it becomes rather significant for the F255W
band, being of the order of $0.5$ magnitudes for $T_{\rm eff}<10,000$~K.

The negligible dependence of the extinction in F439W and F555W 
on effective temperature clearly shows that the sizeable interstellar
reddening affecting both clusters cannot explain the tilted HB
morphology through the effect of $T_{eff}$ dependence of reddening
corrections for the optical bands.

\subsection{Differential reddening}\label{sec:diffredd}

\begin{figure*}
\begin{center}
\includegraphics[width=9cm,height=9cm]{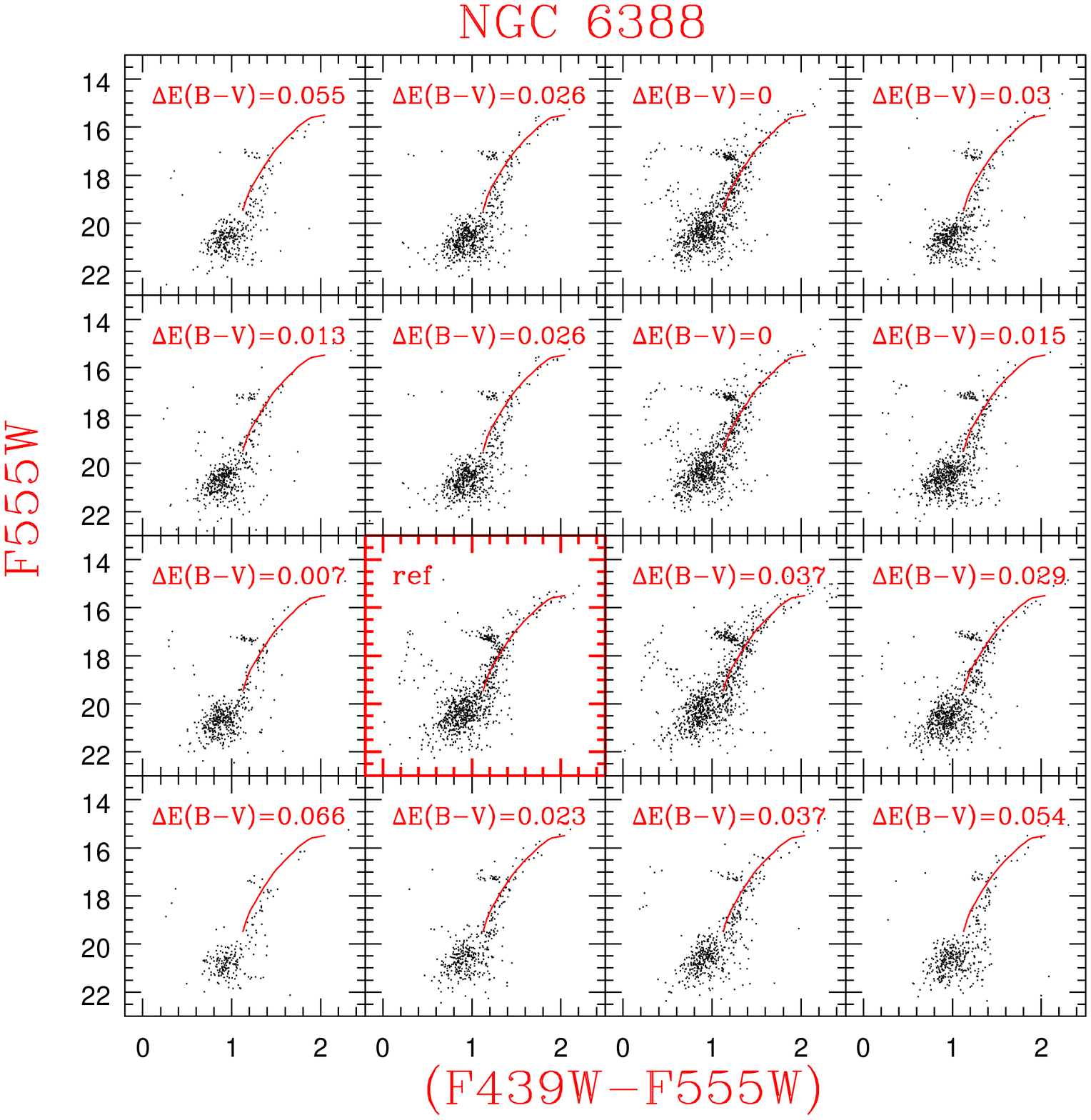}
\includegraphics[width=9cm,height=9cm]{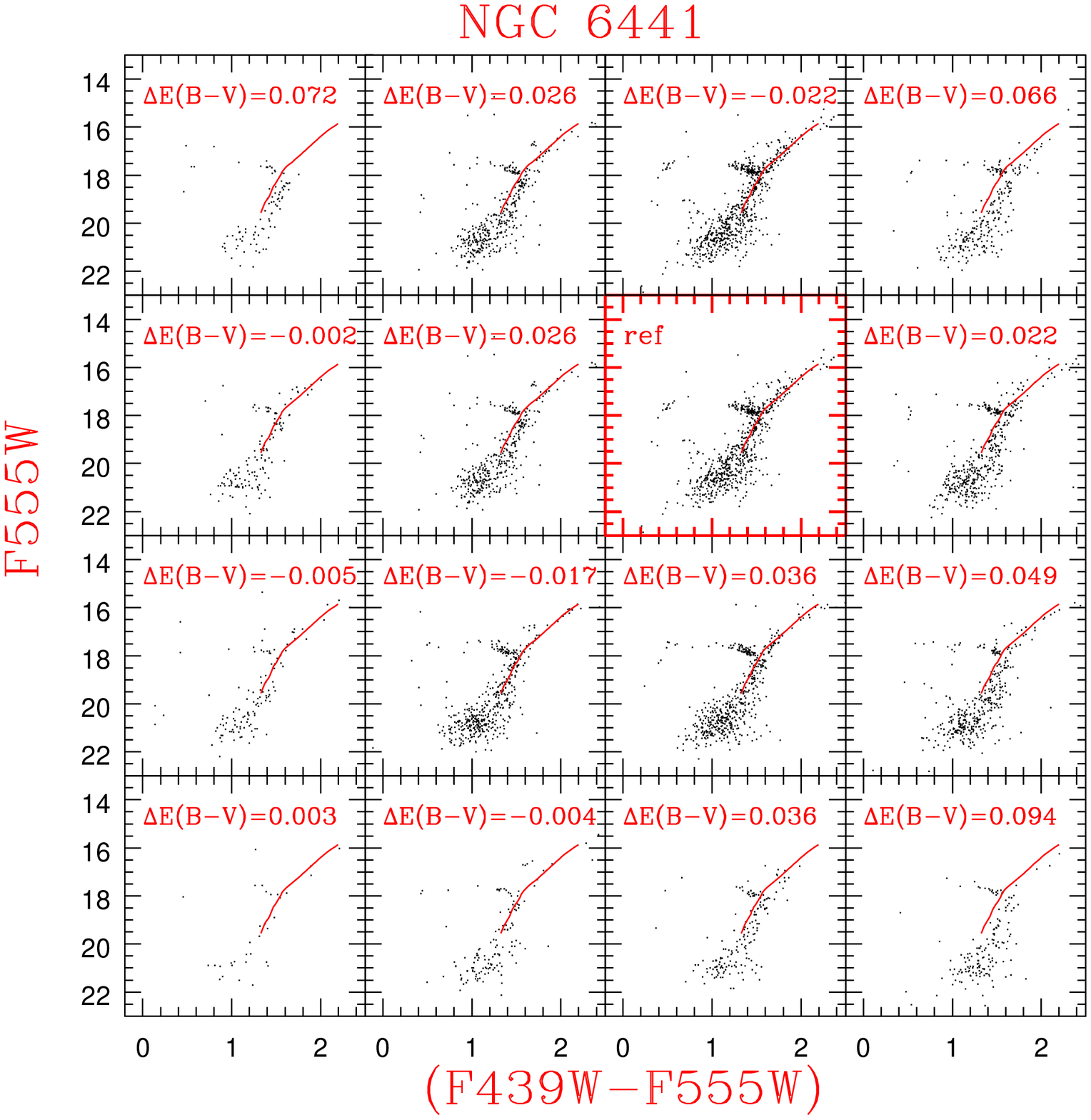}
\end{center}
\caption{Top panel: the CMDs of NGC~6388 from the Planetary Camera field divided into 16 cells of 9 $\times$ 9
arcsec$^2$. These CMDs have been used to estimate the spatial variation of the reddening
in the central region of the cluster. The CMD of the bold panel,
labeled with $ref$, has been arbitrarily adopted as the reference
CMD. The reddening relative to the reference CMD is displayed at the
top left of each region. Bottom panel: same as top panel but for NGC~6441.}
\label{fig:16cmd}
\end{figure*}

In the previous section we have shown that a reddening
correction depending on the stellar effective temperature is not
enough to remove the tilt of the HB. 
However, our approach assumes an average reddening, while it is
well known that both clusters are affected by a sizeable amount of
differential reddening (Piotto et al. 1997, Layden et al. 1999,
Heitsch \& Richtler 1999, Raimondo et al. 2002).  We have applied the
same procedure used by Raimondo et al.~(2002) to both clusters in
order to estimate the size of this effect. We divided the Planetary
Camera field into 16 regions of $9\times$9 arcsec$^2$, and, for each
region, the corresponding CMD has been plotted.  We selected one of
these CMDs as the reference CMD, and extracted its fiducial points by
hand. In Fig.~\ref{fig:16cmd} the CMDs for the various
regions in the two clusters are compared with the fiducial line
obtained by fitting a spline to the fiducial points of the reference
CMD. The average displacement in color of the CMD from the fiducial
sequence is also listed.

The bottom panel of Fig.~\ref{fig:16cmd} is similar to the corresponding figure in Raimondo et
al.~(2002) for NGC~6441, confirming the presence of differential
reddening of the order of 0.10-0.12 magnitudes also on scales of the
order of 10 arcsec. The top panel of Fig.~\ref{fig:16cmd} shows that also NGC~6388 is affected by
differential reddening, though the effect is half as large as in
NGC~6441.  On the other hand, the slope of the tilted HB in the two
clusters is quite similar [$dV/d(B-V)\approx 1.5$], suggesting that differential reddening plays a marginal role in determining the
size of the tilt. Such evidence is further supported by the fact
that the HB of each box of Fig.~\ref{fig:16cmd}  shows a slope consistent
with that observed for the whole sample\footnote{In a few cases this
is not clearly visible, because of the small number of stars in the
selected region.}, also in those regions where the narrowness of the
RGB suggests that no strong residual differential reddening is
present.  In summary, as in Sweigart \& Catelan (1998) and Raimondo et al.~(2002), we are forced to
conclude that the sloped HB is the consequence of an intrinsic
property of the HB stars in NGC~6388 and NGC~6441.

\subsection{The HB morphology in various photometric bands}\label{sec:HBmorph}

\begin{figure}
\begin{center}
\includegraphics[width=8.8cm,height=8.8cm]{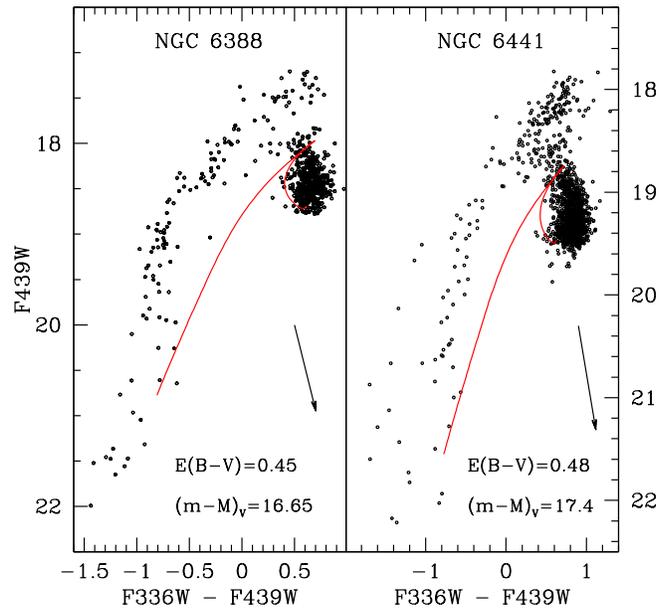}
\end{center}
\caption{(F336W-F439W, F439W) CMD of NGC~6388 and NGC~6441.  The
  reddening and distance modulus values determined by best fitting
 in this diagram  the red clump with the canonical ZAHB models for $Y=0.256$ are labeled. These canonical ZAHB models are plotted by using a solid line. The arrow represents the reddening vector in this photometric plane. The peculiar shape of the ZAHB sequence, the \lq{glitch}\rq\ appearing at color (F336W-F439W)$\sim0.5$, is due to the fact that this color index is not a monotonic function of the stellar effective temperature, coolward of the Balmer maximum at $T_{eff}\approx8500$K.}
 \label{fig:fit_ub}
\end{figure}

In the following discussion, we compare theoretical models
with the various CMDs. We adopt the reddening and distance
modulus values for which ZAHB models with the canonical Y=0.256 best
fit the empirical distribution of the red HB clump in the (F336W-F439W,
F439W) CMD (see Fig.~\ref{fig:fit_ub}), i.e. $E(F439W-F555W)$=0.45 and $(m-M)_{F555W}$=16.65 for NGC~6388, while
for NGC~6441 we adopt $E(F439W-F555W)$=0.48 and $(m-M)_{F555W}$=17.4.

This choice of fitting stellar models to the red clump is clearly arbitrary. In fact, one could in principle try to fit the theoretical predictions to the observations at any point along the HB. This notwithstanding, we consider that the adopted approach for the fitting is reliable for the following reasons: i) in the literature (see the quoted references) devoted to discuss the peculiar features (tilts, jumps, etc.) disclosed by the most recent observations, it is a common procedure to match theoretical ZAHB models to the red HB clump - therefore, for the sake of consistency with previous works we decided to use the same approach; ii) cool ZAHB models should be - at least in principle - more reliable than ZAHB models with an hotter ZAHB location ($T_{eff}$ larger than about 10.000K), since they are not affected by diffusive processes such as atomic diffusion and radiative levitation\footnote{On the other hand, one has to keep in mind that cool HB models - once transferred on the various observational planes - could be affected by larger uncertainties eventually present in the adopted color - effective temperature relations.}.

\begin{figure*}
\begin{center}
\includegraphics[width=12cm,height=11cm]{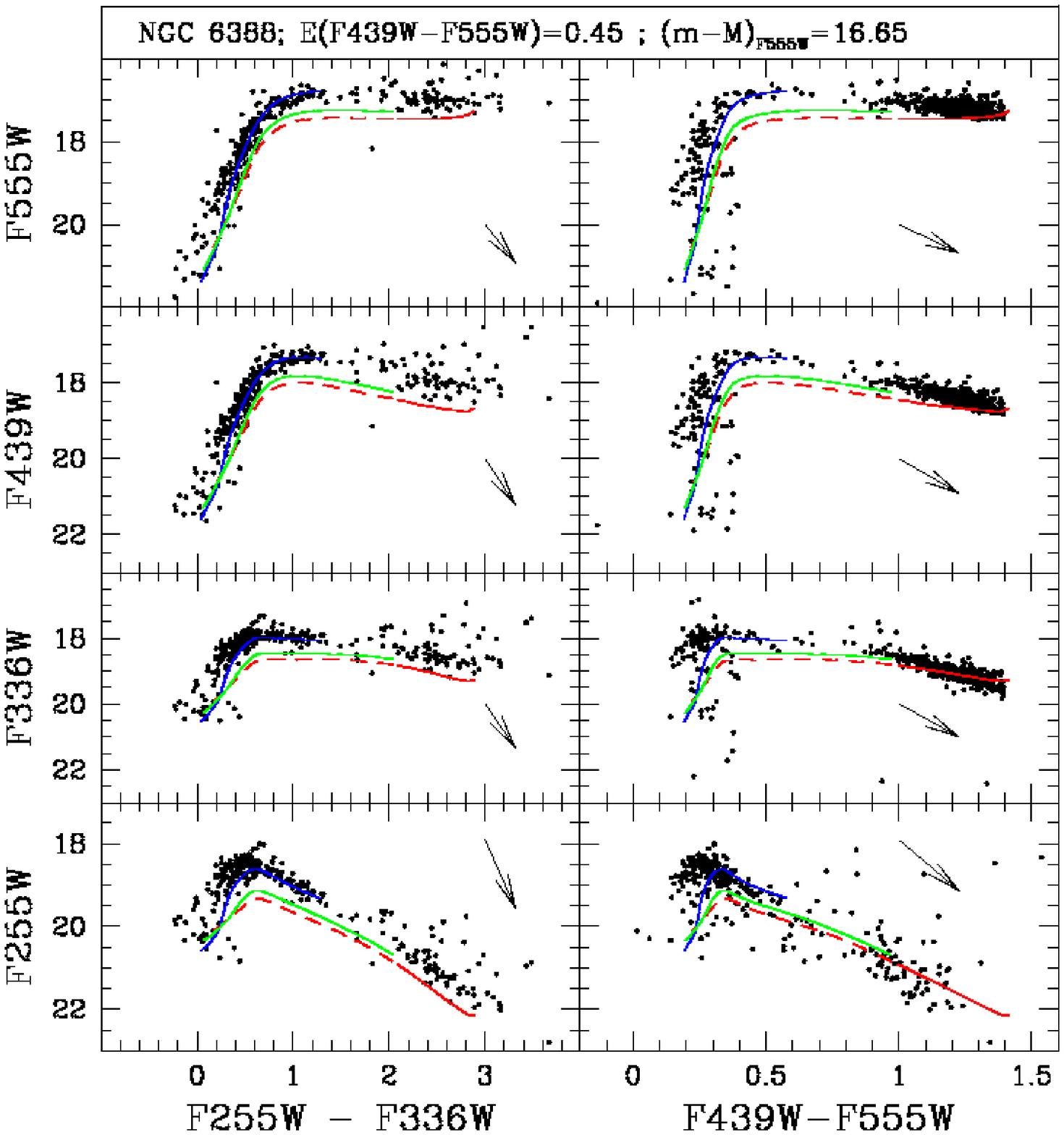}
\includegraphics[width=12cm,height=11cm]{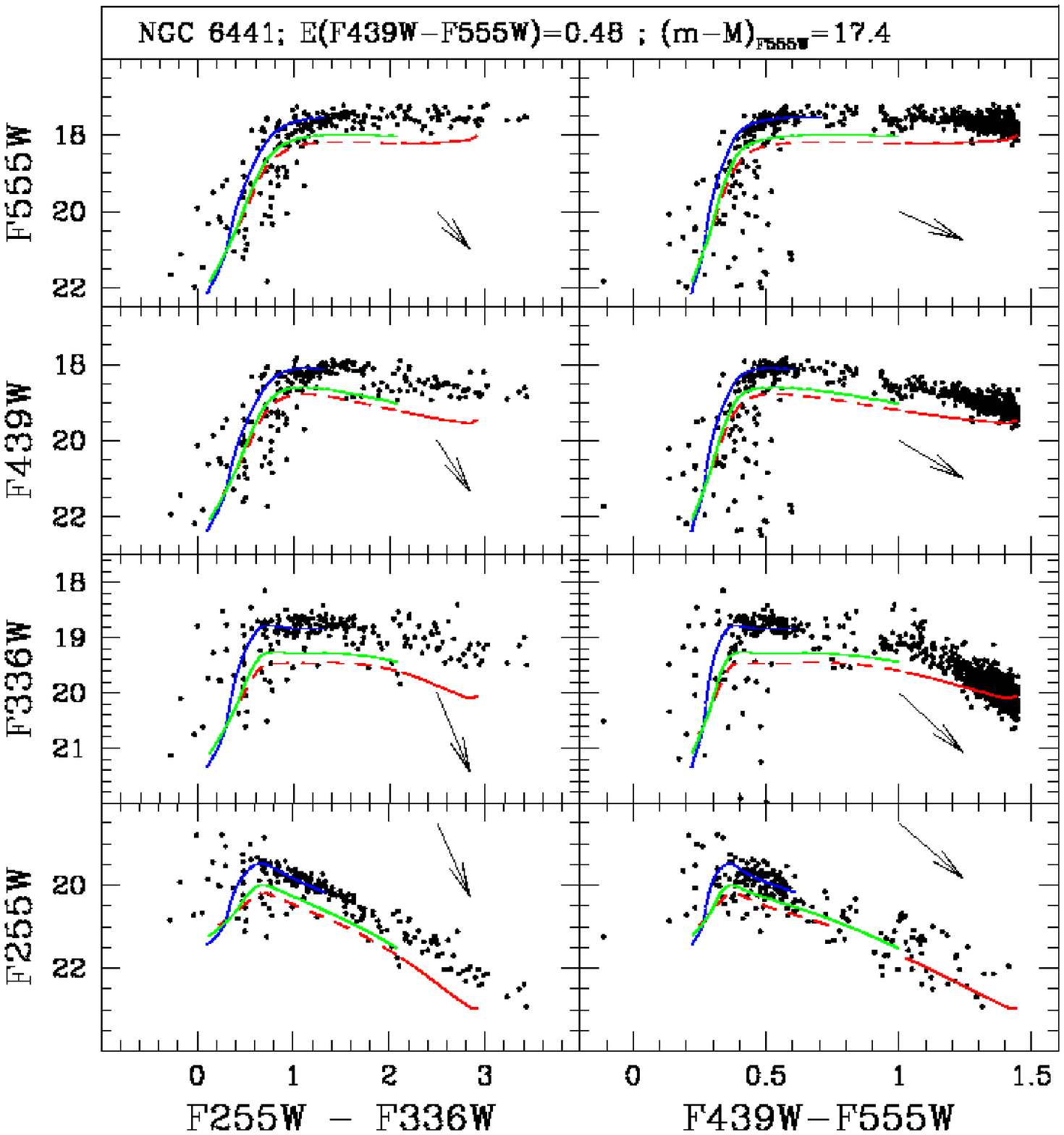}
\end{center}
\caption{Top panel: the HBs of NGC~6388 in different photometric planes. 
   The observations are compared with  theoretical models for a canonical
    Y=0.256 (red line), for Y=0.30 (green line), and for Y=0.40 (blue line). 
The ZAHB models account, in all cases, for a
reddening correction dependent on $T_{\rm eff}$ (see text for
details). The shifts applied to the stellar  models for the adopted distance modulus and mean reddening are labeled
in the figure, and they have been fixed according to our (arbitrary)
choice of imposing a best fit between models and the lower envelope of
the red HB portion in the (F336W-F439W, F439W) CMD (see Fig.~\ref{fig:fit_ub}). The arrows represent the
reddening vector in the different photometric planes.  Bottom panel: as in the top panel, but for NGC~6441.}
\label{fig:diff_cmd}
\end{figure*}

Figure~\ref{fig:diff_cmd} shows a comparison, in the various photometric planes,
between the ZAHB models and the HB sequences of NGC~6388 and NGC~6441. 
For the following discussion we refer only to the HB models with the canonical helium abundance Y=0.256, plotted in the CMD with a red line.
The comparison between theory and observations clearly
shows that it is not possible to have an overall fit of the entire
HB: if we impose to the models that they fit the red HB, there is no way they can
fit the blue part, and vice versa. The striking evidence is that this
is true for all combinations of colors and magnitudes, which raises
the suspicion that there must be a real mismatch between the canonical
ZAHB models and the observed HBs for both clusters.  
We wish to note that, so far, this is the first clear empirical evidence showing that the tilted morphology of the HB of the two target GCs is present also in the UV photometric bands -  which are much better suited than optical bands for studying hot HB stars.

In passing, we note also that it is not possible to find a unique combination of
distance modulus and reddening that allows to properly fit the HB in
the various observational planes, even if we consider only the red
part of the HB.
This occurrence is quite evident in the (F439W-F555W, F255W) CMD, and it
is likely due to the combination of differential reddening and of the fact
that for the red portion of the HB we start to lose stars (because
they are too faint) when using the F255W magnitudes.

As already mentioned, when analyzing the optical plane (F439W-F555W, F555W), a tilt is evident when moving
from the red HB clump toward bluer colors (see also Rich et al.~1997,
Sweigart \& Catelan~1998 and Raimondo et al.~2002), with stars close to the HB turnover
which are 
$\approx0.5$ magnitudes brighter than the average
magnitude of the red HB. It is worth noting once again that this tilt
is present in all photometric planes, and it appears even more evident
in the UV CMDs, where the blue part of the HB is up to
$\approx 1$ magnitudes brighter than the red part.
  
It appears -- at least in some CMDs, such as the (F336W-F439W, F439W) one in Fig.~5 --
that the red portion of the theoretical ZAHB is not able to approach
the reddest colors of the observed red HB. Again, this occurrence
could be due either to some problems in the adopted color-$T_{\rm eff}$
relation, or to an effect of the differential reddening affecting
these clusters (see the discussion about the change of the red clumpy
HB morphology as a consequence of differential reddening made by
Raimondo et al. 2002), or to a combination of both effects.

\section{The blue--hook star candidates}\label{sec:bluehook}

The data shown in Fig.~\ref{fig:diff_cmd} reveal that 
the models 
with canonical Y=0.256 are not able to reproduce the peculiar
morphology of the HB in both clusters. Although this is the most
evident shortcoming of the models, a careful examination of the same figures
shows also that the models are not able to reproduce the color of the
hottest HB stars in the CMD. This empirical finding could be
interpreted as evidence that the effective temperature of these
stars is significantly higher than that of the hottest HB models
(whose total mass is virtually coincident with the He core mass)
that can be envisaged on the basis of canonical stellar evolution
theory.

This occurrence is evident for NGC~6388, but much less clear in
NGC~6441, which is not surprising, since we already noted that the latter
has a much less populated and less extended blue HB. The extension of
the observed HB beyond the theoretical one is visible in almost all
the CMDs of NGC~6388 (Fig.~\ref{fig:diff_cmd}), in particular in all
the UV CMDs, which are more appropriate for the study of the hottest
stars. Again, this mismatch could be an artificial effect of the
transformation from the theoretical to the observational plane and/or
be due to differential reddening. However, the presence of a possible gap
(see Sec.~\ref{sec:cmds_counts}), just beyond the location of the
hottest stars in the canonical models suggests that the stars
observed beyond the gap could have a different nature. Therefore it is
tempting to associate them to the blue hook stars already observed by
D'Cruz et al.~(2000) in $\omega$~Centauri (NGC~5139) using far-UV data,
and by Brown et al.~(2001) in NGC~2808, by means the far-UV and near-UV
cameras of the Hubble Space Telescope Imaging Spectrograph.

Although Brown et al.~(2001) and Cassisi et al.~(2003) have
shown that the observational properties of these objects cannot be
understood in the framework of canonical stellar evolution theory, we
rely on canonical ZAHB models in order to obtain 
a rough
estimate of their effective temperature: they appear to
have a $T_{\rm eff}$ larger than 
about 25,000~K. It is worth noticing that such an effective temperature
limit appears significantly lower than the estimated minimum effective temperature
(see Brown et al. 2001 and D'Cruz et al. 2000) for blue hook stars.
The present analysis does not allow us to verify if this difference
is due to the quite approximate approach used for estimating the
temperature of these stars, 
or whether it reflects a real intrinsic difference between blue hook stars
in metal-intermediate clusters such as NGC~2808 or $\omega$~Cen and metal-rich ones as our
target clusters.
The left panels of Fig.~\ref{fig:box} of the present paper provide visible evidence supporting
the first detection  of this feature in metal-rich stellar systems. We tentatively
identify as blue-hook candidates all those stars we plot as asterisks.

The blue-hook extension is different in the two clusters: in NGC~6388
the sample of blue hook star candidates is
much larger, and its location in the CMD is quite better defined than in the
case of NGC~6441.  In NGC~6388, the stars belonging to the blue hook
show on average increasing color when moving towards fainter
magnitudes, while in NGC~6441 the hottest part of the CMD is more
dispersed and, on average, the hottest stars have a similar color as
the brighter and cooler stars belonging to the same sequence.
This occurrence could be due, in part at least, to the larger
absolute and differential reddening affecting this cluster.
Another possibility to explain the significant color spread in the candidate blue hook stars
could be related to the physical mechanism which has been suggested
to be at the origin of these peculiar objects, i.e. the so-called delayed He Flash
scenario discussed in the following.

As noticed by many authors since Rich et al.~(1997), the existence of
EHB stars in NGC~6388 and NGC~6441 is quite difficult to be accounted
for in the framework of canonical stellar evolution theory, which
predicts that the He flash (HeF) takes place at the tip of the Red
Giant Branch (RGB). None of these stars would have a hot enough
location on the HB able to explain the presence of extended blue tails
in any Galactic GCs, though, as recently shown by Castellani et
al.~(2006), this problem could be less severe than previously claimed.
Castellani \& Castellani (1993) suggested the so-called delayed
HeF scenario in order to explain the existence of blue tails. This
scenario envisages that, as a consequence of a high mass-loss
efficiency during the RGB evolution, due to enhanced stellar winds
and/or dynamical interactions within dense cluster cores, a stellar
structure can lose such a large amount of envelope mass that it fails
to ignite the HeF at the RGB tip, being forced to evolve toward the
White Dwarfs (WDs) cooling sequence. Depending on the amount of
residual envelope mass, it will still be able to ignite He-burning
either at the bright end of the WD cooling sequence (the so-called
\lq{early}\rq\ hot flasher, EHF) or along the WD sequence (the
so-called \lq{late}\rq\ hot flasher, LHF). After the HeF, the star
will settle on its ZAHB location, but due to the low
mass of the remaining H-rich envelope, its ZAHB location will be
significantly hotter than that of canonical stellar models. The first
attempt to find a connection between the scenario outlined by
Castellani \& Castellani for EHF stars (see also D'Cruz et al.~1996) and the presence of the
blue hook was made by D'Cruz et al.~(2000) 
in $\omega$ Centauri, though they found that the
predicted ZAHB location of EHFs is not hot enough to explain the empirical
evidence.

This scenario has been analyzed also by Sweigart (1997) and more
recently by Brown et al.~(2001), who devoted particular attention
to the evolution of LHFs: this preliminary analysis showed that the
convection zone arising with the late HeF, which occurs under
conditions of extreme electron degeneracy, is able to penetrate into
the H-rich envelope, causing the mixing of H into the hot He-burning
regions (HeF-induced mixing) where it is burnt quickly. A successive dredge-up
process of matter that has been processed via both H- and He-burning
would enrich the stellar envelope in helium, carbon and nitrogen.
According to the quoted authors, these surface abundance changes  would cause a discontinuous increase of the
effective temperature along the HB at the transition between unmixed
(EHF) and mixed (LHF) models, thus producing a gap at the hot end of the
HB as indeed observed by Bedin et al.~(2000) in NGC~2808, and which
could be tentatively related to the possible gap at $F555W\approx
19.8$ in the HB of NGC~6388.  

From an observational point of view,
further support to this scenario was provided by the accurate
spectroscopical measurements made by Moehler et al.~(2002, 2004) for
the blue-hook stars belonging to NGC~2808 and $\omega$~Cen. The most
important finding was the evidence that blue-hook stars are both
hotter and more helium-rich than predicted by
canonical models of EHB stars. In recent times, the results of Brown
et al.~(2001) have been fully confirmed on the basis of
full self-consistent evolutionary computations performed by Cassisi et
al.~(2003).

The changes in the surface chemical composition in the LHF structures
are really crucial for explaining the fact that the blue-hook stars appear
fainter than the \lq{normal}\rq\ hot HB stars. Indeed, LHF stars, as
the EHFs, do have smaller He cores at the HeF with respect to
canonical models, but the expected smaller core mass implies a
reduction in the bolometric luminosity of only $\approx0.1$
magnitudes.  However, it has been nicely shown by Brown et al.~(2001)
that LHFs have significantly different spectra with respect to the other stars
belonging to the same clusters, but which have not experienced the HeF
mixing process. In a normal stellar atmosphere composed mostly of H,
the opacity for wavelengths shortward of about 910
{\AA} redistributes the flux in the extreme ultraviolet to longer
wavelengths. If He is enhanced in the stellar atmosphere as in LHF
stars, the H opacity source is strongly reduced and so much more
flux is emitted in the extreme UV, decreasing the outgoing flux at
longer wavelengths.  As a consequence, LHF stars should appear fainter
in the far and intermediate UV bands (as F255W) compared to 
EHF and normal HB stars.

It is worth noting that, depending on the properties of the HeF-induced mixing
and on the efficiency of the following dredge up, one can expect non-negligible differences 
in the surface abundances in the sense that some LHF stars might appear more enriched in helium and
carbon than others. This occurrence could result in a sizeable spread both in
brightness and colors among the LHF stars (as appears to happen in the case
of the blue hook candidates in the target clusters).

The empirical data we present for NGC~6388 and NGC~6441 nicely
confirm these predictions, and reproduce similar findings in NGC~2808
(Bedin et al.~2000, Brown et al.~2001, Castellani et al.~2006) and
$\omega$~Cen (D'Cruz et al.~2000, see also Momany et al.~2004). What
is interesting to note in the case of NGC~6388 (for NGC~6441, the
number of stars does not allow a firm conclusion) is that the
brightness extension of the blue-hook sequence seems to be larger than in the other
quoted clusters. It is not possible to clarify, with the present data,
whether this occurrence is a signature of a real intrinsic difference
between the blue hook stars belonging to the different clusters, or if it
is due to the different photometric bands adopted in the various
works (see the discussion in Brown et al.~2001), or to
photometric errors/contamination by foreground stars, or more simply
to higher differential reddening in NGC~6388. In any
case, it is evident that an accurate spectroscopic investigation of the
blue-hook stars in -- at least -- NGC~6388, as already done for NGC~2808
and $\omega$ Cen (Moehler et al.~2002), is mandatory. Moehler and Sweigart
(2006b) have tried to spectroscopically measure at least the temperature
and gravity of the hottest stars in NGC~6388, but they had to conclude
that crowding might be a severe obstacle for ground-based spectroscopy. 
Since HB stars are rare in the cluster outskirts, and mixed
with bulge objects, a selection of cluster stars (possibly based
on proper motions on wide-field images as in Anderson et al.~2006) would 
be of great help in this kind of investigation.

\section{The tilted HBs}

Another intriguing property of the HBs of NGC~6441 and NGC~6388 is the occurrence of
the tilt for which we still lack a convincing explanation. In the
following, we wish to briefly summarize the state-of-art of the
investigations on this interesting feature, and provide further
evidence which could help in interpreting it.

On a purely empirical ground, Raimondo et al.~(2002) have shown that a
tilted red HB is a common occurrence among all metal-rich GCs. This
feature appears quite more evident in NGC~6388 and NGC~6441 since
these clusters are the only ones showing a blue extension of their HB. 
On the other hand, Catelan et al.~(2006) seem to find that the red HB is more strongly sloped in NGC~6388 than in 47~Tuc over the same color range (their Fig.~3).
The most simple explanation for the occurrence of such a feature is to
assume that two stellar populations, characterized by distinct metal
abundances, exist in these clusters. According to this scenario the blue
HB stellar sample would be associated with the metal-poor population,
while the red clump would be produced by the metal-rich component.

However, this possibility has been completely ruled out by both the
analysis performed by Raimondo et al.~(2002), who showed that in both
clusters there is not a corresponding sizeable number of RGB stars
blueward of their RGB (see also Catelan et al.~2006), and by the spectroscopic measurements of
individual metal abundances of RGB stars performed by
Carretta (private comunication) and Gratton et al.~(2006).  In NGC~6388,
Carretta has found an average metallicity of
[Fe/H]=$-0.44$, with no evidence of a metallicity spread among the 7 RGB stars for which high-resolution UVES@VLT spectra have
been obtained.  This result has also been recently confirmed by \cite{walle07}.
Again, from high-resolution UVES@VLT spectra for
5 NGC~6441 red giants, Gratton et al. found an average
metallicity [Fe/H]=$-0.39$, with no indication of a metallicity spread
larger than their measurement errors. On the other hand,
it must be mentioned that Clementini et al.~(2005) have found that out
of 11 RR Lyrae in NGC~6441, 9 objects have a remarkably similar metal content
[Fe/H]$=-0.57$, whereas 2 objects have a metallicity
[Fe/H]$\approx-1.3$, though, in this case, the membership of the RR Lyrae
is not certain.  In view of the results by Gratton and collaborators,
a more accurate analysis of the RR Lyrae membership is necessary in
order to obtain a firmer result. In any case, also a metallicity as
low as [Fe/H]$=-1.3$ could not explain the presence of an EHB in NGC~6441.

Piotto et al.~(1997) suggested the possibility that in NGC~6441
and NGC~6388 there are two populations with different ages: the oldest
population should be responsible for the blue HB, while the younger
one creates the red HB. This possibility would indeed explain the
bimodal HB, but it would not be able to account for the tilted HB.

Moreover, quite recently Catelan et al.~(2006) provided an accurate
CMD for NGC~6388 reaching the cluster main-sequence turn off. On the
basis of their photometry, they conclude that the age of this cluster
is quite similar (within 1Gyr) to that of the prototype metal-rich cluster
47~Tuc, with no clear evidence for the existence of an age spread in
NGC~6388.

An attempt to simultaneously interpret the occurrence of the blue
tail and of the tilt was made by Sweigart \& Catelan (1998) by
suggesting three different non-canonical scenarios: i) a very high
initial He content, ii) a spread in the He core mass as due to stellar
rotation, iii) deep mixing along the RGB. Each one of these mechanisms
can -- in principle -- explain the presence of both the blue tail and
the tilt. However, as shown by Raimondo et al.~(2002), Moehler and Sweigart (2006b) and Catelan et al.~2006, these non-canonical scenario seem to
be  ruled out by some clear empirical evidence. The primordial high He
content is excluded by (i) the luminosity of the bump along the RGB,, and (ii) the ratio between
the number of HB and RGB stars -- the so-called R parameter -- which are both 
consistent with a canonical He content (Layden et al. 1999, Cassisi et
al. 2003, Salaris et al. 2004).
Fast rotation is an unlike possibility because, in all GCs with blue HBs, no stars
hotter than $T_{\rm eff}\sim11,500$K have a rotation velocity $v \sin~i
> 10$ km/s   (Recio Blanco et al.~2002). However, Sweigart~(2002) has pointed out
that blue HB stars could appear as slow rotators as a consequence of
an efficient removal of angular momentum from the envelope due to stellar
winds induced by radiative levitation. Clearly, more reliable HB stellar models
properly accounting for diffusive processes, mass loss and rotation are mandatory
in order to fully exploit this issue. In the same time,  it would be worthwhile to investigate the properties
of hot HB stars via the asteroseismological analysis of those HB stars experiencing non-radial pulsations such as EC~14026 objects and \lq{Betsy}\rq\ stars (see  Kawaler \& Hostler~2005).

Finally, deep mixing along the RGB, often invoked to explain the
abundances anomalies observed in cluster RGB stars, has been recently seriously
questioned from an observational point of view by the identification of the same anomalies among main-
sequence stars (see Gratton, Sneden, and Carretta 2004 for a
discussion), which points towards a primordial origin for the abundance
anomalies. Nevertheless, in more recent times, Suda \& Fujimoto~(2006) have discussed
the possibility that RGB stars in Galactic GCs could be affected by various mechanisms of mixing between the envelope and 
the interiors. These different modes of mixing would be triggered by close encounters with other stars that RGB stars
could experience in the quite dense stellar environment characteristic of many GCs. The simulations made by Suda \& Fuijmoto~(2006)
have suggested that depending on the occurrence of a specific mode of mixing as well as the evolutionary stage of the
RGB star experiencing the process, the HB progeny of these stars could populate the hot side of the HB and also
at a luminosity level significantly larger than that predicted by canonical stellar models.

\subsection{A possible solution to the enigma}

An interesting scenario that could help to solve the puzzle of the
extended and tilted HBs of NGC~6388 and NGC~6441, has recently emerged
from some quite unexpected results that some of us obtained for the
Galactic GCs $\omega$~Cen and NGC~2808. As already suspected by Anderson
(1997), Bedin et al.~(2004) showed that the main sequence (MS) of
$\omega$~Cen is splitted into two sequences, and that this double MS  
is a ubiquitous feature, spread all over the cluster. Even more
interestingly, Piotto et al.~(2005), by using high-resolution spectra,
showed that the bluer MS is more metal-rich than the red one, at odds
with any expectation based on stellar evolution theory. The only
possibility to account for a bluer and more metal-rich secondary MS
(which contains about 25\% of $\omega$~Cen´s MS stars) is to assume (as
already proposed by Bedin et al.~2004 and Norris 2004) that it
corresponds to a highly helium-enhanced (Y=0.38) stellar
population. Such a population must necessarily reflect a second generation
of stars formed from material polluted by intermediate-mass AGB stars,
and/or by type II supernovae, and/or massive rotating stars (Caloi \& D'Antona 2005, Maeder \&
Meynet 2005, see also discussion in the Piotto et al. 2005, and Bekki \&
Norris 2006).  Since He-rich stars have a lower turn-off mass at a given
age, they will populate the blue side of the HB during the He-burning
phase, and they will also be more luminous because of the higher
energy produced by the hydrogen-burning shell (Lee et al. 2005 and references therein). Both effects would
produce a bluer and brighter HB, exactly as observed in NGC~6388 and
NGC~6441.

The HB of $\omega$~Cen is too complicated (because of the large spread
in metallicity and age) to clearly distinguish the effect of He
enhancement in a minor component of its stellar population from the more
prominent effects due to age and metal content. Moreover,
$\omega$~Cen may well be an extreme example in which ejecta by a
first generation of massive stars can be kept inside a Galactic GC,
because of its mass. Indeed, $\omega$ Cen is the most massive GC
in our Galaxy, and it might well have been much more massive in the past as,
because of its orbit, it must have lost a huge amount of mass due to
tidal shocks, mainly passing through the bulge (see Altmann et al.~2005 and reference therein). 

However, Piotto et al.~(2007) have clearly shown that the $\omega$ Cen
MS split is not a unique case among GCs: also NGC~2808 shows three
distinct main sequences. An anomalous broadening of the MS of NGC~2808
was already indicated by D'Antona et al.~(2005).
The simplest explanation of this
occurrence is that the various sequences are associated to stellar
populations characterized by different initial He
contents
(D'Antona et al. 2005, Piotto et al. 2007). Interestingly enough,
NGC~2808 is the Galactic GC with 
one of the most complex HB structures: a red HB clump, a very extended
blue tail and very few RR Lyrae stars 
(Sosin et al.~1997, Bedin et al.~2000, Corwin et al.~2004). One has also to note that
NGC~2808, is the most massive GC after $\omega$~Cen.

Once again, NGC~6388 and NGC~6441 are between the most massive GCs
and, because of their location inside the Galactic bulge, they must
have been rather more massive in the past \cite{ree02}, in order to
have managed to survive till now. Indeed, according to Gnedin and
Ostriker (1997), both clusters have a disruption time slightly shorter
than a Hubble time, which means that what we see at the present time
might be the remnants of much bigger stellar systems.  On the other
hand, the existence of populations with an enhanced He content has
been proposed recently \cite{Kaviraj07}
also in the case of extragalactic GCs, suggesting that this phenomenon is typical not only
of our galaxy but is in fact more general.

It is therefore rather tempting to associate the anomalously blue and
anomalously tilted HBs of NGC~6388 and NGC~6441 to a second generation
of stars, strongly He-enriched by pollution from massive and/or intermediate-mass stars of the
first star formation burst. 
A similar suggestion that the presence of a population with He
enhancement can explain the anomalous HB of NGC 6388 and NGC 6441 has
been recently made also by Caloi and D'Antona (2007).  
Note that this
scenario is completely different from that early envisaged by
Sweigart and Catelan (1998), i.e. that the HB morphology of NGC~6388
and NGC~6441 may be due to a population of He-enriched stars. Sweigart and
Catelan suggested that all stars in these two clusters are He-rich. 
Now we suggest that a minority of the stellar
population of NGC~6388 and NGC~6441 is He-enriched.

In order to verify the possibility that the peculiar HB morphology
could be explained by accounting for the presence in the target clusters
of a He-rich stellar population, we have calculated additional sets
of low-mass, He-burning models for suitable choices of the initial
He content.  More in detail, we computed HB models for a metallicity
Z=0.008 and inital He contents equal to Y=0.30 and 0.40. The mass of
the RGB progenitors ($\sim 0.7M_\odot$) has been selected in order to
fulfill the condition that their age at the RGB tip be equal to
about 13~Gyr. It is worth noticing that these models are fully
consistent with the other ones computed by adopting Y=0.256, as far as
both the physical input and numerical assumptions are concerned. The
stellar models have been transferred from the theoretical plane to the
various observational ones by using the same set of color-$T_{\rm
eff}$ relations adopted for the Y=0.256 models.
It has been recently shown by Girardi et al. (2007) that the impact
of an enhanced He abundance on color-$T_{\rm eff}$ relations is completely
negligible.

In Fig.~\ref{fig:diff_cmd}, we show a comparison between the observed CMDs and
the theoretical models computed for the various He contents, adopting
the same reddenings and distance moduli as explained in Sec.~\ref{sec:HBmorph}. These
reddenings and distance moduli have been fixed by imposing that the
canonical models overlap the red clump of the HB in the optical
bands. As already discussed, the models computed by assuming a canonical He content (Y=0.256) are
not able to reproduce the blue HB. On the contrary, one can easily see
that the ZAHB models corresponding to Y=0.40 are in good agreement
with the observed distribution of blue HB stars of NGC~6388 in all of
the CMDs (top of Fig.~\ref{fig:diff_cmd}).  The same kind of comparison, but for the case of
NGC~6441, is performed in the bottom panels of Fig.~\ref{fig:diff_cmd}. In this case, it is evident that the
Y=0.40 ZAHB is slightly brighter than the observed distribution of
blue HB stars.  From the comparison between empirical data and the ZAHBs
computed for various initial He contents, it appears that it is
possible to reproduce the brightness of the blue HB star population
by assuming a He content of the order of $Y\approx0.35$, i.e.
slightly lower than in NGC~6388. This fact is consistent with the
observational evidence that the blue HB is less extended in NGC~6441
than in NGC~6388.

The presence of two distinct stellar populations characterized by two
different initial He contents can help in explaining the brightness
difference between the red portion of the HB and the blue
component. However, in order to explain the tilted morphology of the
whole HB sequence one should also account for the presence of a spread
in the He content at the level of about $\Delta{Y}\approx0.05--
0.06$ (see also Moehler \& Sweigart 2006b,
Caloi \& D'Antona 2007). 

\begin{figure}
\begin{center}
\includegraphics[width=9cm,height=9cm]{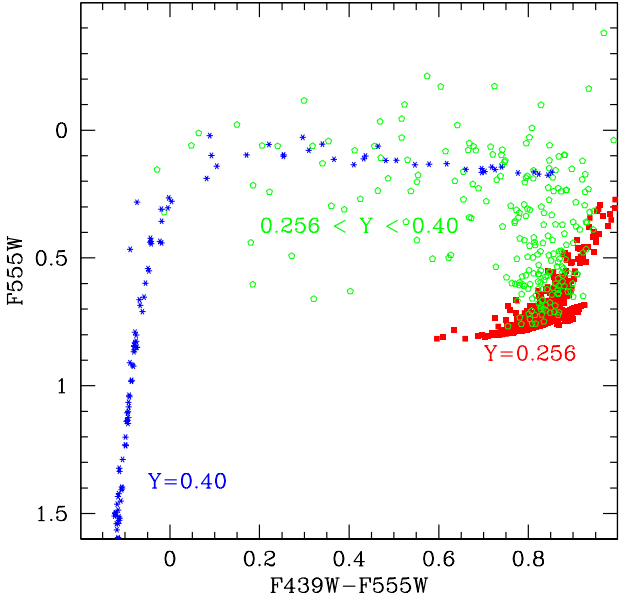}
\caption{Synthetic HB simulation for NGC~6388 (see text for more details). The various colors refer to HB structures with 
different initial He content, namely: red color for stars with Y=0.256 and blue color for those with He equal to 0.40,
green points correspond to the HB structures whose He abundance is the range $0.256 <Y< 0.40$.}
\end{center}
\label{fig:hb_He}
\end{figure}

In passing, we note that the presence of a moderate spread ($\Delta{Y} \approx 0.02-0.03$) 
in the He abundance around the canonical value ($Y\sim0.26$) among the red clump stars, 
by affecting their intrinsic luminosity, could help 
explain the peculiar tilted distribution of the red HB clump, too.

In order to provide further support to the hypothesis that the peculiar HB morphology in NGC~6388 and
NGC~6441 could be explained as due to the existence in both clusters of stellar populations characterized by
different He contents, and eventually to estimate the fraction of stars belonging to the He-enhanced
stellar population, we performed an analysis based on the computation of synthetic HBs by adopting
stellar model sets for different initial He abundances: Y=0.256, 0.35 and 0.40.

\begin{figure}
\begin{center}
\includegraphics[width=9cm,height=9cm]{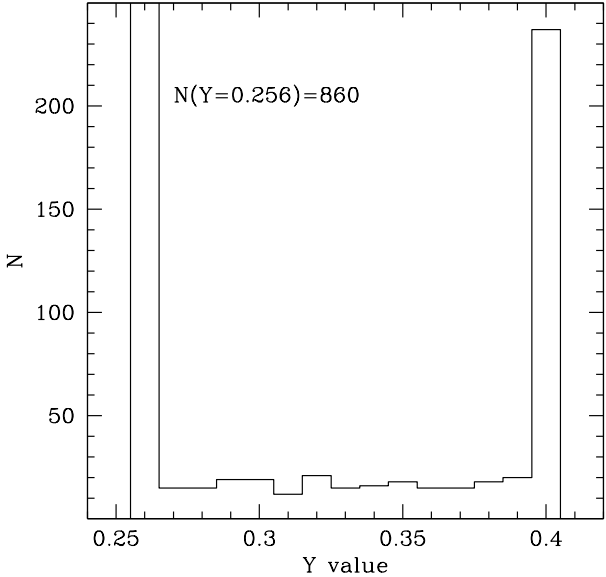}
\caption{The number distribution of HB objects with the initial He content, corresponding to the synthetic HB simulation showed in the
previous figure. The total number of objects corresponding to the \lq{canonical}\rq\ He content $N(Y=0.256)$ is also labeled.}
\end{center}
\label{fig:numbery}
\end{figure}

The details of the computation of synthetic HBs are fully discussed by Castellani, Caputo \& Castellani (2003).
In our synthetic HB simulation, we assume that the cluster members are a mixture of stars with different initial He
abundances\footnote{We note that this approach is quite similar to that adopted by Caloi \& D'Antona (2006)
for reproducing the HB morphology of NGC~6441.}. More in detail, our synthetic HB simulations have been performed
by adopting the following scheme:

\begin{itemize}

\item We first generate a HB stellar generation corresponding to a \lq{canonical}\rq\ He abundance Y=0.256, by adopting a
Gaussian distribution with a mean mass $M_{HB}^{0.256}$ and dispersion equal to $\sigma=0.02M_\odot$;
\item A second stellar generation is computed by assuming an initial He content Y=0.40, mean mass $M_{HB}^{0.40}$ and same
sigma as above;
\item The existence of a stellar population with an initial He abundance between the canonical value and Y=0.40 is accounted
for by assuming a uniform He content distribution (between Y=0.256 and 0.40), a Gaussian mass distribution
with mean value $M_{HB}^{-Y-}$ and sigma equal to $0.02M_\odot$.
For each He value in the quoted range, the corresponding HB evolutionary tracks used in the synthetic HB simulations have been obtained by linear interpolation between the HB models computed for the previously quoted initial He abundances.
\end{itemize}

In performing the various simulations we tried to reproduce various observational constraints such as the number of stars
along the different portions of the HB (see Table 2) and the number of RR
Lyrae stars and their mean pulsational period. 
It must be clearly stated that,
due to the evident large number of free parameters in the simulation,
it is 
not appropriate to attribute too much 
physical relevance to these simulations\footnote{As early noticed by
  Caloi \& D'Antona~(2007), the number of RR Lyrae stars
as well as their mean pulsational period are largely dependent not only on the choices made for the synthetic simulation
but also on the random number extraction.},
which should be used just to give a general, qualitative picture of
the scenario. We think that it is more realistic to consider them as a
useful tool to roughly estimate the number of stars in the various
stellar populations -- for the different initial He contents -- that
reproduce well the quoted empirical constraints.

In Fig.~\ref{fig:hb_He}, we show the result of an HB simulation that
satisfactorily reproduces the shape\footnote{The present simulations do not
account for photometric errors as well as for differential
reddening.}, as well as the star counts, of the HB in NGC~6388. This
simulation has been performed by adopting:
$M_{HB}^{0.256}=0.65M_\odot$, $M_{HB}^{0.40}=0.55M_\odot$ and
$M_{HB}^{-Y-}=0.65M_\odot$. 
It provides 220 objects in the blue side
of the HB, 25 RR Lyrae and 1061 
stars in the red side of
the HB. These numbers appear in good agreement with those reported in
Table~3. The distribution of stars with the initial He content is
shown in Fig.~8: there are about 860 objects in the peak corresponding
to the canonical He content, 230 
stars for a He content equal to
Y=0.40, and 240 objects whose He content is in the range 0.256 - 0.40.
The mean RR Lyrae pulsational period we obtain from this simulation is equal to
$P_{RR}\sim0.85 d$.

The previous analysis has shown that it would be possible to explain
the various observational peculiarities of the HBs in NGC~6388 and
NGC~6441 by accounting for the presence in both clusters of distinct
 stellar populations characterized by two different 
initial He contents: the canonical value $Y\sim0.26$ and a quite large abundance $Y\sim0.40$; 
plus an additional stellar population whose initial He content is -- more or less uniformly --
spread between the two extreme values.

\begin{figure*}
\begin{center}
\includegraphics[width=16cm,height=8cm]{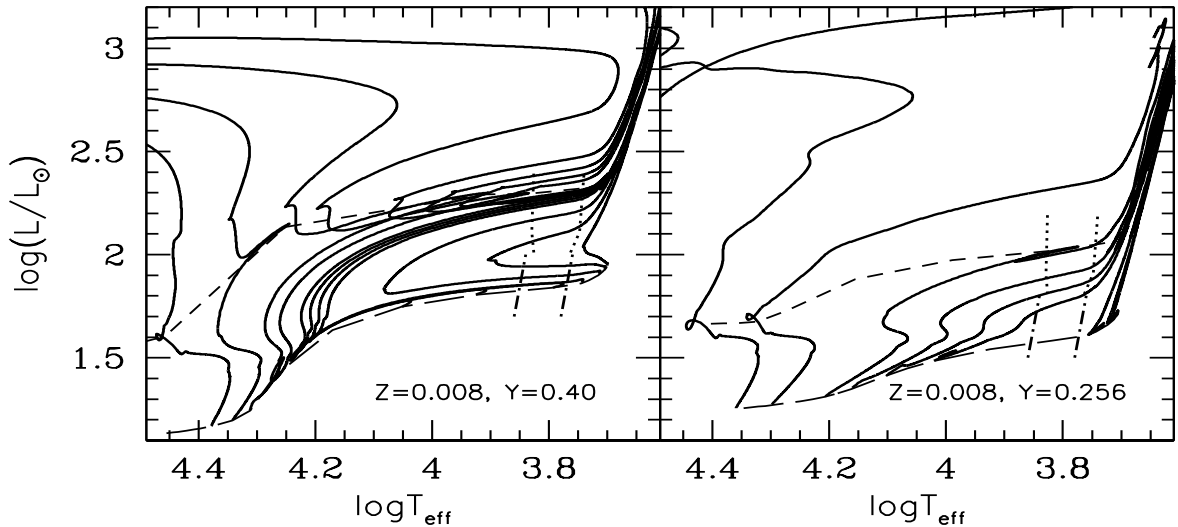}
\caption{Left panel: the H-R diagram for selected low-mass, He-burning
models for Z=0.008 and Y=0.40. The stellar masses are (in solar unit):
0.47, 0.48, 0.49, 0.50, 0.52, 0.53, 0.54, 0.55, 0.60, 0.65, 0.70. The
ZAHB locus (long dashed line) and the central He-exhaustion one (short
dashed line) are shown. The blue and red edges of the RR Lyrae
(dot-dashed line) and Type II Cepheids instability strip (dotted line)
are also plotted. The luminosity of the instability strip
boundaries have been - arbitrarily - increased by 0.2 dex in order to
account for the larger He content. Right panel: as left panel but for
Y=0.256. In this case the evolutionary tracks for the 0.47 and
0.48$M_\odot$ models are not present. The luminosity levels of
the instability strips shown are those reported in literature.}
\end{center}
\label{fig:theo_y}
\end{figure*}

 At the beginning of this section, we
noted that the clearest, empirical support to the existence in a
GC of multiple stellar populations with different He content is the
splitting of the MS observed in the clusters $\omega$ Cen and
NGC~2808. Therefore, one could argue if the same circumstantial
evidence could be observed in the two target clusters. 
We really think that it is extremely challenging to obtain a similar result 
for the two target clusters for the following reasons: 
1) they are both affected by a sizeable amount of
differential reddening, 2) the full interpretation of the HB
morphology would require the presence of some amount of spread in the
initial He content between the different stellar populations.  
Both processes affect the CMD location of MS stars with the global effect
of making it a thorny problem to detect any splitting of the MS locus.
In any case, it would be extremely worthwhile to face this difficult task.

\section{Summary and discussion}

We have presented multiband optical and UV CMDs of the Galactic GCs
NGC~6388 and NGC~6441. We focused our attention on their
anomalous HBs, which show a number of similarities, but also some
interesting differences.
 
The results of the analysis carried out in the previous sections can
be summarized as follows:
\begin{itemize}

\item The main peculiarities of the HBs are: (i) their extension to very
  hot temperatures, including (at least for NGC~6388) a number of blue
  hook candidates, which probably correspond to late helium flasher stars; (ii)
  the presence of a tilt in the HB;\\

\item The peculiar HBs of NGC~6388 and NGC~6441 cannot be reproduced
  by any canonical models;\\

\item Reddening and differential reddening
  contribute to create a sloped HB, but neither of these effects is
  sufficient to explain the observed tilted HB;\\

\item The presence of a He-rich stellar component -- with initial He
content in mass $Y$ larger than 0.35 -- allows to explain the observational
evidence of a blue HB, and the fact that the blue HB in both clusters
is brighter than the red HB clump. The presence of a He spread between the
\lq{canonical}\rq\ value and $Y\sim0.35 - 0.40$ allows to explain the observed
upward slope of the HB in both clusters;\\

\item A comparison with theoretical models computed by adopting
various initial He contents suggests that in both NGC~6388 and NGC~6441 
the difference in brightness between the red and blue sides of the HB is consistent 
with the presence in the cluster of
a stellar component with a He content as high as
$Y\approx0.40$ in NGC~6388 and $Y\approx0.35$ in NGC~6441. In addition, in order to explain the whole tilted morphology of the HBs in these clusters, one has to invoke the presence of an additional stellar population characterized by a spread -- between the two extreme values -- of the initial He content.
On the basis of a suitable simulation of synthetic HBs for NGC~6388,
we estimate that the more He-enhanced stellar population in this cluster should contribute to
about 18\%  of the total stellar population\footnote{On the basis of the HB star counts, we estimate that the fraction of He rich stars in NGC~6441 should be lower, i.e. of the order of 10\%.}. Similarly,
the stellar population with initial He abundance in the range $0.26<Y<0.40$ should contribute about 16-17\% of the global population;\\

\item The tilted morphology of the red part of the HBs could also be explained as due to the combined
effects of: differential reddening, existence of a (small) He spread among the red clump stars;\\

\item We think that a firmer interpretation of all of the empirical
  findings presented in the current paper can be achieved only when
  the reliability of the color - $T_{\rm eff}$ relation and the bolometric
  correction scale for metal-rich, cool stars is carefully investigated, as well as when a firmer understanding of the processes regulating the efficiency of mass loss along the RGB is achieved.\\

\end{itemize}

In the framework of the scenario proposed in this work, the presence
of a He-enriched population implies that in both clusters there must
have been at least two episodes of star formation, with the second
generation of stars polluted by material ejected from the first formed
stars: for the possible culprits, see the scenario described by Ventura et al.~(2001, 2002) for GCs in general and 
the discussion for the case of $\omega$ Cen in Piotto et al.~(2005), Bekki and Norris (2006) and Karakas et al. (2006). The main stars responsible for the pollution
could be Type II supernovae, rotating massive stars, and
intermediate-mass AGB stars. Of course, it remains to be understood how
the ejected material can remain inside the potential well of the
cluster (and it appears not enriched in metals as it should be if the He-enriched matter is provided by Type II supernovae). 
This problem is beyond the purposes of the present
paper. However, it is important to note that, after $\omega$ Cen and NGC~2808,
NGC~6388 and NGC~6441 are amongst the most massive clusters of the
Galaxy. Because of their position within the bulge, it is conceivable
that what we observe at the present time are just two remnants of much
bigger systems, which lost part (most?) of their stellar populations
via successive tidal shocks, mainly with the Galactic bulge, as
expected also from their disruption time, smaller than a Hubble time.

By using synthetic HB models, we have already shown that the presence of multiple
stellar populations with different initial He contents could account for many empirical
facts related to the peculiar morphology of the HB in both clusters. However, as already noticed,
in these simulations there are many free parameters such as the mean HB mass and the mass dispersion
for each stellar population, as well as the adopted distribution law for the initial He abundance. This drastically reduces the predictive strenght of this approach. Therefore in order to
rely on \lq{first principles}\rq, we now wish to briefly comment on the
differences in the evolutionary tracks corresponding to HB stellar
structures with different initial He contents in order to show how the morphology
of these tracks alone is able to provide an explanation for some empirical findings. 
In Fig.~9, we show
the evolutionary tracks in the H-R diagram of selected HB models with
two different initial He contents, Y=0.256 and Y=0.40, i.e. the
same values adopted in the previous analysis. The selected stellar
masses are the same in both panels, with the only difference that for
Y=0.40 two additional hot HB models have been plotted. In both panels,
we show also the ZAHB locus and the locus corresponding to the central
He-exhaustion (CHeEX). The boundaries of the RR Lyrae (Bono et al.~1997, the same adopted for
the synthetic HB simulations) 
and type II Cepheid (Di Criscienzo et al.~2007) instability
strips are also plotted.

There is an obvious, expected difference in the brightness of the
evolutionary tracks due to the difference in the envelope He content
and, in turn, in the shell H-burning efficiency. An interesting
feature is that, in the case of the Y=0.40 models, the luminosity range
between the ZAHB and the CHeEX at the level of the instability strip
is about 25\% larger than in the case of the Y=0.256 models. This
occurrence has the important consequence that, if one assumes the same
mass distribution along the HB for a He-rich and for a He-normal
stellar population, the He-rich one has a higher probability to
produce brighter and, in turn, longer-period variables.  An
additional interesting feature is the evidence that, due to the
peculiar morphology of the He-rich stellar tracks (once again related
to the high efficiency of the shell H-burning), many stellar models
densely populate a region of the H-R diagram, within the instability
strip, at luminosities around $\log{(L/L_\odot)}\approx 2.25$. It is
important to notice that this occurs before the end of the core
He-burning phase, i.e., when their evolutionary lifetimes are not too
rapid, and therefore there is a no negligible probability to observe them.
All these morphological properties clearly tend to increase the
chance, for a He-rich old stellar population, to produce brighter,
long-period, variables. 

By comparing the morphology of the evolutionary tracks for the
different initial He contents with the predicted boundaries of both
the RR Lyrae and Type II Cepheids instability strip, it appears clear
that a He-rich stellar population has a probability to produce a
larger number of Type II Cepheids than a normal He stellar population:
this might provide a simple explanation for the presence in
both NGC~6388 and NGC~6441 of a large population of Type
II Cepheids.

A firmer understanding of this issue would require additional
simulations of synthetic HBs.
However, as already noticed, in these simulations
the number of free parameters is quite large and this significantly hampers the reliability
of this kind of investigation. In a forthcoming paper, we will check if the analysis of the various types of variable stars
as well as their pulsational properties and period distribution can help to better constrain the
scenario outlined in the previous section. 

In section 4.3, we discussed the reasons for which we decided to match the observed CMD and the models at the red end of the HB distribution. In the following, we wish to very briefly comment some implications of using a different choice for matching theoretical predictions to the empirical data, as for instance to force the fit at a point blueward of the red HB clump. When using this approach, one obtains that the red HB clump is definitively fainter than canonical models and this occurs also for the hottest HB stars\footnote{This occurrence could be not a real problem when considering the possible presence of blue hook stars.}. 

None of the evolutionary scenarios outlined in section 6 for explaining the HB morphology of NGC~6388 and NGC~6441 is able to predict a fainter location for the red clump, nor can this occurrence be explained as due to (both absolute and differential) reddening on the basis of the evidence previously discussed. If we accept the hypothesis of multiple stellar populations characterized by different initial He content, a fainter location for the red HB clump would require the presence of a stellar population with an He content lower than the canonical one, a possibility that appears highly unreliable if not impossible.

In addition, Catelan et al. (2006), once forced the red HB clump of NGC~6388 to match the HB of 47~Tuc, have found that the two clusters are approximately coeval. Therefore, we are now facing with two possibilities:
\begin{itemize}
\item  we could assume that the red HB stars of 47~Tuc are affected by some anomalies as those of NGC~6388 - but this appears at odds with all empirical evidence - and then the result obtained by Catelan et al. (2006) is still valid; 

\item otherwise we could assume that the HB stars in 47~Tuc are \lq{normal}\rq\ and those in NGC~6388 are affected by some problem,  and then they should appear fainter than those in 47~Tuc. However, if this is the case, the age of NGC~6388 should be larger of about $\sim5-6$Gyr with respect that of 47~Tuc. This result appears to us quite unreliable. 

\end{itemize}

On the basis of these considerations, we are confident that the approach adopted in present work, as well as in many others in literature, is reliable.

\begin{acknowledgements}
We warmly thanks our anonymous referee for her/his pertinent comments and useful suggestions.
This work has been partially supported by INAF and MUR. 
GB gratefully acknowledges support from the Deutsche Forschungsgemeinschaft through grant Mo 602/8 and partial support from the ESO Director General Discretionary Fund.. 
SC warmly acknowledge the Instituto Astrofisico de Canarias (Tenerife) for the hospitality he experienced during his stay at IAC.
MC acknowledges support by Proyecto FONDECYT Regular No. 1071002. 
We warmly thank W.~B. Landsman and S. Moehler for detailed readings of the manuscript and for useful comments.
\end{acknowledgements}

\end{document}